\documentclass[conference]{IEEEtran}
\usepackage{cite}
\usepackage{amsmath,amssymb,amsfonts}
\usepackage{algorithmic}
\usepackage{graphicx}
\usepackage{textcomp}
\usepackage{xcolor}
\usepackage[hyphens]{url}

\usepackage{subcaption}
\usepackage{pifont}
\usepackage{multirow}
\usepackage{enumitem}
\usepackage{diagbox}

\usepackage[linesnumbered,ruled,lined, noend]{algorithm2e}

\newcommand\blfootnote[1]{%
	\begingroup
	\renewcommand\thefootnote{}\footnote{#1}%
	\addtocounter{footnote}{-1}%
	\endgroup
}

\newcommand{\myparagraph}[1]{\noindent {\bf #1}}

\newcommand{\name}{\textsc{\mbox{BlackJack}}~}
\newcommand{\areaoverhead}{2.46\%~} 
\newcommand{\pwroverhead}{3.28\%~} 
\newcommand{\latoverhead}{0.56\%~}

\def\BibTeX{{\rm B\kern-.05em{\sc i\kern-.025em b}\kern-.08em
		T\kern-.1667em\lower.7ex\hbox{E}\kern-.125emX}}

\title{\name: Secure machine learning on IoT devices through hardware-based shuffling}
\author{
	\IEEEauthorblockN{Karthik Ganesan, Michal Fishkin$^*$, Ourong Lin$^*$, Natalie Enright Jerger}
	\IEEEauthorblockA{University of Toronto
		\\\{karthik.ganesan, michal.fishkin, hughorlin\}@mail.utoronto.ca, enright@ece.utoronto.ca}
}

\begin{document}
\maketitle
\thispagestyle{plain}
\pagestyle{plain}

\begin{abstract}		
Neural networks are seeing increased use in diverse Internet of Things (IoT) applications such as healthcare, smart homes and industrial monitoring~\cite{mohammadi2018}.
Their widespread use makes neural networks a lucrative target for theft. 
An attacker can obtain a model without having access to the training data or incurring the cost of training. 
Also, networks trained using private data (e.g., medical records) can reveal information about this data~\cite{fredrikson2014}. 
Networks can be stolen by leveraging side channels such as power traces of the IoT device when it is running the network.
Existing attacks require operations to occur in the same order each time; an attacker must collect and analyze several traces of the device to steal the network. 
Therefore, to prevent this type of attack, we \textit{randomly shuffle} the order of operations each time.
With shuffling, each operation can now happen at many different points in each execution, making the attack intractable.
However, we show that shuffling in software can leak information which can be used to subvert this solution. 
Therefore, to perform secure shuffling and reduce latency, we present \name, hardware added as a functional unit within the CPU.  
\name~secures neural networks on IoT devices by increasing the time needed for an attack to centuries, while adding just \areaoverhead area, \pwroverhead power and \latoverhead latency overhead on an ARM M0+ SoC.\blfootnote{$^*$ Both authors contributed equally to this work.}	
\end{abstract}
	
\section{Introduction}
\label{s:introduction}
\vspace{1em}

The Internet of Things (IoT) has enabled novel applications in fields such as health monitoring~\cite{hassanalieragh2015}, smart homes~\cite{samuel2016} and remote sensing~\cite{pallavi2017}. 
Within IoT, Machine learning (ML) is seeing increased use in areas such as image and voice recognition, indoor localization and biomedical monitoring~\cite{mohammadi2018}. 
With their increasingly widespread deployment, ML models have become appealing targets for theft.
There are many reasons why an attacker would wish to steal a ML model deployed on an IoT device:
\begin{itemize}[leftmargin=*, noitemsep]
	\item Classification accuracy is highly dependent on access to high-quality training data, which an attacker might not have access to. 
	Obtaining a pre-trained model obviates the need for this training data, allowing an attacker to replicate the accuracy of a well-trained model. 
	\item A trained model can leak information about the training data, which must remain confidential. 
	For models trained using patient medical records for example, leaking training information can result in a serious breach of privacy~\cite{fredrikson2014}.		
	\item In networks used for financial applications, reverse engineering the network would allow an attacker to bypass fraud detection.  
	For example, EMVCo (i.e., `chip and pin'), used by Visa and MasterCard, employs neural networks for fraud detection~\cite{ouerdi2015}.
	An attacker with access to this model could learn how to circumvent fraud detection and charge credit cards without getting caught.		
\end{itemize}	
For these reasons, it is critical that ML models deployed on IoT devices be secured against attackers. 

Direct access to the models stored in on-chip memory is normally blocked by manufacturers.
For example, TI's MSP430FR chips require a password to access the JTAG port~\cite{ti-jtag}. 
However, attackers can still use side channels to gather secret information from the device. 
Side channels are vectors such as timing, power consumption or electromagnetic emanations (EM) which can leak information about data being processed by the device~\cite{meneghello2019}.
Prior work shows that side-channel leakage can be used to fully reverse engineer a neural network running on an IoT device~\cite{Batina2019, Maji2021, Joud2023}. 
By analyzing EM traces of the device running the network, an attacker can learn the size, activation function and the weights for every layer.\footnote{We elaborate on the full details of the attack in Section~\ref{s:attack}.}
While these attacks target neural networks, we show that they can also apply to other ML algorithms, such as autoencoders and support vector machines (Section~\ref{s:attack-extending}).

Power side-channel attacks require collecting and analyzing several traces, to eliminate the effect of noise from other system components. 
For this analysis to work properly, the operations being targeted must occur in the same place in each trace. 
Therefore, one approach to thwart such attacks is to \textit{randomly shuffle} the order of operations each time.
When shuffling is applied to neural network layers, each weight is used at a different point for every inference run. 
The attacker must then try every possible combination of the recovered weights to carry out a successful attack.
For example, to reverse engineer $M$ shuffled weights, the attacker would need $\mathcal{O}(M!)$ traces to mount an attack.
For a single neuron with $64$ weights, shuffling increases the number of traces needed for a successful attack by \emph{90 orders of magnitude}.
If an attacker collects and analyzes $1000$ traces a second, they would need $4.026\times10^{78}$ years to reverse engineer the weights of a single neuron.
In \name, we also shuffle the order of neurons ($N$) per layer.\footnote{As we describe in Section~\ref{s:solution}, we shuffle the order of iterations for convolutional and pooling layers as well.}
Thus, the total number of possible permutations increases to $\mathcal{O}(M!)\times\mathcal{O}(N!)$.
As neural networks consist of hundreds of neurons and thousands of weights, collecting enough traces to reverse engineer a whole network would take millions of years, making the attack completely untenable.
For example, in the networks we evaluate in Section~\ref{s:system-evaluation}, the largest values of $M$ and $N$ are $5670$ and $128$.

Prior work has shown that shuffling is effective at preventing side-channel attacks targeting neural networks~\cite{Brosch2022, Nozaki2021}.
However, these works implement shuffling in software, which suffers from a number of drawbacks:
\begin{enumerate}[leftmargin=*, noitemsep]
	\item Software shuffling leaks side-channel information. We demonstrate a new attack which undermines the security benefits of software shuffling (Section~\ref{s:software-shuffling}).
	\item Software shuffling adds significant latency overheads due to the additional CPU instructions required (Section~\ref{s:evaluation}).
\end{enumerate}
To overcome the limitations of software shuffling, we propose \name, hardware to perform random shuffling.
\name is added as a functional unit within the CPU, which significantly reduces the latency overhead of shuffling (Section~\ref{s:system-evaluation}).
While prior work has proposed hardware for shuffling, these designs are limited to shuffling $2^N$ objects~\cite{Dhanuskodi2019, Chen2023, Bayrak2012}.
This limitation makes existing approaches unsuitable for shuffling the arbitrary number of weights and neurons used in neural networks.
\name provides an efficient, low-latency hardware solution which supports shuffling any number of values.
Furthermore, \name is `symmetric' (i.e., it does not leak information based on the current input) and therefore does not leak any side-channel information (Section~\ref{s:evaluation-coco}).
While we focus on securing neural networks deployed on IoT devices, \name can also be used to secure other applications which operate on sensitive data (Section~\ref{s:discussion-otherapps}).
Finally, we show that \name can also thwart other side-channel attacks against neural networks, such as floating-point timing attacks and fault-injection attacks (Section~\ref{s:evaluation-otherattacks}). 

In summary, we make the following contributions:
\begin{itemize}[leftmargin=*, noitemsep]
	\item We show that shuffling is an effective technique to prevent side-channel attacks against ML algorithms, due to the large number of operations that can be shuffled.
	\item To the best of our knowledge, we show the first side-channel attack against software shuffling, to learn the exact values being shuffled.
	An attacker can use this information to `undo' shuffling and carry out the attack as before. 
	\item To perform shuffling securely and with much less overhead, we add \name as a functional unit within the CPU. 
	\name effectively prevents side-channel attacks, while adding just \areaoverhead area, \pwroverhead power and \latoverhead latency overhead to an ARM M0+ SoC.
	\item We demonstrate the versatility of our approach by showing that \name is effective at preventing other side-channel attacks as well as securing other applications against such attacks. 
\end{itemize}

\section{Background and Related Work}
\label{s:background}

In this section, we provide background on side-channel attacks and prior work on shuffling, a commonly used technique to defend against these attacks.

\subsection{Side-channel attacks}
\label{s:attack-background}

Side-channels attacks are a widely used mechanism to obtain secret information about a system, without interfering with normal system operation.
Attacks such as SPECTRE~\cite{spectre} and MELTDOWN~\cite{meltdown}, which target large out-of-order cores, have highlighted the strength of side-channel attacks. 
SPECTRE and MELTDOWN use \textit{timing} side channels, where an attacker leverages the time difference between certain operations to steal secret information. 
As IoT systems typically employ very simple processors, they are more commonly targeted by power side channel attacks~\cite{meneghello2019}.

Performing a power side-channel attack requires collecting traces of the system being targeted. 
A trace is a measurement of the device while it is operating on secret data.
The power trace varies based on the secret information being operated on by the device.
Thus, by analyzing these power traces, an attacker can reverse engineer the secret information used.
To collect these traces, an attacker only needs access to the voltage ($V_{dd}$) input of the device.
A commonly used proxy for power is to measure the Electromagnetic (EM) emanations of the device.
This does not even require the attacker to physically contact the device at all; the EM probe must simply be placed near the device~\cite{meneghello2019}. 

A key difference between timing and power/EM side channels is the number of traces required; with timing channels, information can be leaked with a single trace. 
However, for power/EM attacks, many traces are required to recover secret information.
This is because these side channels are noisy due to interference from other system operations~\cite{meneghello2019}.
Thus a single trace does not provide sufficient resolution for an attacker to recover information.
The attacker must therefore collect a large number of traces and analyze them together to eliminate noise.
Thus, variations between the traces makes the attack more difficult as the attacker must compensate for variations before performing the attack.
One popular technique for preventing side-channels attacks is \textit{masking}, which we describe in Section~\ref{s:related-masking}.
We now focus on the other common technique, \textit{shuffling}, which is the basis of our work. 

\subsection{Shuffling}

Shuffling randomly reorders the sequence of sensitive operations each time a program is run~\cite{mangard2007}.
With operations happening at different points in each trace, the attacker can no longer identify the position of each operation.
Therefore, shuffling $N$ operations forces the attacker to collect $N!$ traces to account for every possible ordering. 

We provide a detailed survey of prior approaches which employ shuffling in Section~\ref{s:related}.
Our work differs from prior approaches for hardware shuffling in two major ways: 1)~We target neural networks, which have hundreds of neurons and thousands of weights. 
The major limitation of shuffling for securing AES is that there are only $16$ S-Box values to shuffle, which limits the number of possible permutations to $16!$ for AES.
Thus, our use of shuffling for securing neural networks results in a huge number of possible permutations, and consequently the time needed for a successful attack tremendously.
2)~Unlike prior works which can only shuffle the order of $M$ operations when $M$ is a power of 2, \name can efficiently shuffle the order of operations for any value $N$. 
To the best of our knowledge, \name is the first technique to perform hardware shuffling for arbitrary values of $M$.

Randomly shuffling operations requires a means to secure produce random numbers.
For this purpose, we use a True Random Number Generator (TRNG), a hardware module to produce a sequence of random bits.  
TRNGs use some physical phenomenon (e.g., power supply noise, temperature, voltage fluctuations) to generate random numbers~\cite{Sunar2009}.
By relying on such analog phenomenon, TRNG outputs do not conform to a repeating pattern than an attacker can learn to subvert the security of the TRNG.
On supported systems, the TRNG output can be accessed in software using a random number generation function (e.g., $rand()$ in C).
We assume that both software shuffling and \name use a TRNG for generating random numbers.

\section{Attacking neural networks}
\label{s:attack}

In this section, we explain how the neural network running on an IoT device can be stolen via side-channel attacks. 
We focus on power/EM side-channel attacks and describe other types of attacks in Section~\ref{s:evaluation-otherattacks}.
While several power/EM side-channel attacks have been proposed~\cite{Batina2019, He2022, Joud2023}, we focus on CSI NN~\cite{Batina2019} as a representative side-channel attack from this class.
We then describe how we replicate the CSI NN attack.
Finally, we show how this attack can be extended to ML models other than neural networks.

\begin{figure}[t]
	\centering
	\includegraphics[width=0.8\columnwidth]{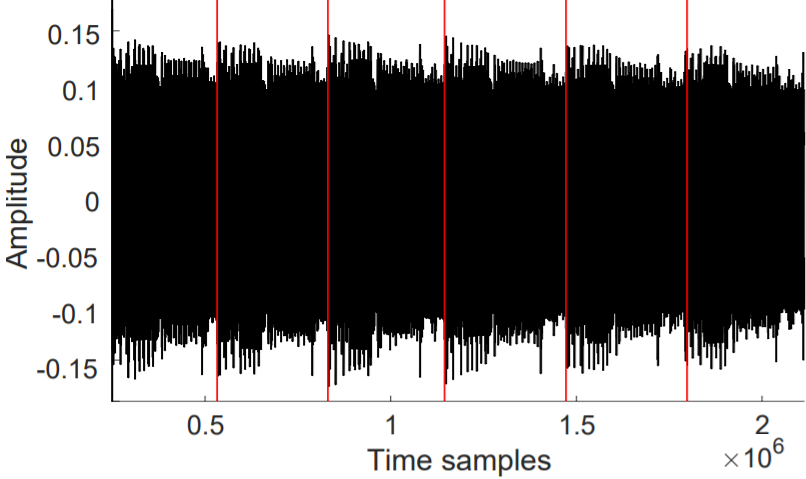}
	\caption{Identifying the number of neurons in a layer~\cite{Batina2019}.}
	\label{figure:csi-neuron}
\end{figure}

\begin{figure}[t]
	\centering
	\includegraphics[width=0.8\columnwidth]{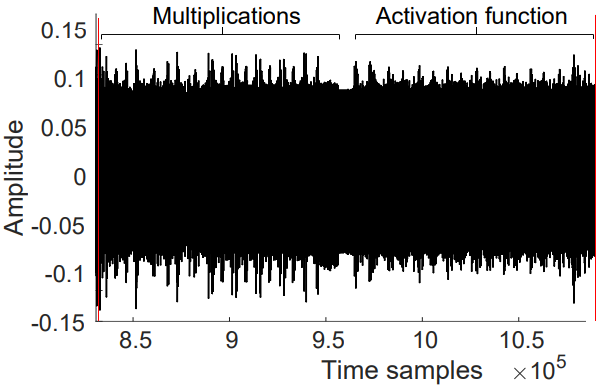}
	\caption{Identifying the number of weights per neuron and the activation function used per layer~\cite{Batina2019}. }
	\label{figure:csi-weight}
\end{figure}

\begin{table}
	\small
	\centering
	\caption{Comparison of delays (in ms) for commonly used activation functions running on an ARM M3 CPU~\cite{Batina2019}}
	\label{table:activations}
	\renewcommand{\arraystretch}{1.2}
	\begin{tabular}{|c|ccc|}
		\hline
		Activation function & Min   & Max   & Mean   	\\ \hline
		ReLU                & 5.8   & 6.06  & 5.9  	\\ 
		Sigmoid             & 152 	& 222 	& 189 	\\ 
		Tanh                & 51  	& 210 	& 184 	\\ 
		Softmax				& 724	& 877	& 813	\\ \hline
	\end{tabular}
\end{table}

\subsection{CSI NN}

CSI NN uses electromagnetic emanations from an IoT device running a neural network to learn the weights and the hyperparameters (i.e., number of layers, number of neurons per layer and the activation functions) of the network.  
We show how each of these is determined, starting with the network hyperparameters.

\myparagraph{Number of neurons.}
Calculating the output of a neuron consists of several multiplication operations followed by the activation function. 
Figure~\ref{figure:csi-neuron} shows the EM trace for a layer with six neurons.
An attacker needs to simply count the number of neurons from the trace.

\myparagraph{Activation function.}
In Figure~\ref{figure:csi-neuron}, two distinct regions can be seen per neuron. 
The per neuron trace in Figure~\ref{figure:csi-weight} shows several multiplication operations followed by the activation function.
CSI NN observes that common activation functions (i.e., $sigmoid$, $tanh$, $ReLU$ and $softmax$) show significant variations in runtime (Table~\ref{table:activations}).
$ReLU$ takes $<10$ms, while $sigmoid$ and $tanh$ take $50$-$200$ms and $sigmoid$ takes $700$-$900$ms.
This variation can be used to identify the specific activation function used for each layer. 
With the activation function known, the attacker can then split the trace into segments containing only the weights for the next step.

\myparagraph{Weights.}
\label{s:attack-weights}
The next step is to determine the values of these weights using Correlation Power Analysis (CPA), which requires an accurate model of the device's power consumption. 
The power model is highly dependent on the hardware being targeted. 
In microcontrollers, the memory bus consumes the most power~\cite{veyrat2012}.
The memory bus is pre-charged to all 0's before any memory is read.
Then, based on the value read, the power consumed is proportional to the number of bus lines that are charged to 1.
This is known as the \textit{Hamming Weight (HW)} power model and is the most commonly used model for microcontrollers~\cite{Batina2019, Maji2021}.
The attacker then generates `weight candidates' -- a list of all possible weight values and their Hamming weights.

\myparagraph{Correlating traces.}
For this step, the attacker first splits each trace into per-weight segments and targets each weight separately. 
For each weight, the attacker has $D$ power traces (i.e., $t$), each consisting of $T$ measured data points. 
The attacker also has a list of $I$ weight values ($h$), one for each trace (since each trace uses a different input).
Now, the attacker must correlate the measured traces $t$ against the guesses of the power model $h$. 
The Pearson correlation coefficient (PCC) is the most widely used metric for this purpose~\cite{veyrat2012}. 
The PCC ($ \rho $) is calculated using Equation~\ref{e:pearson}.
\begin{gather}
\rho_{t,h} = \frac{\sum_{d}^{D}[(h_{d,i}-\overline{h_i})(t_{d,j}-\overline{t_j})]}{\sqrt{\sum_{d}^{D}(h_{d,i}-\overline{h_i})^2\sum_{d}^{D}(t_{d,j}-\overline{t_j})^2}}
\label{e:pearson}
\end{gather}
The parameters in Equation~\ref{e:pearson} are:
\begin{itemize}[leftmargin=*, noitemsep]
	\item $t_{d,j}$ is point $j$ in trace $d$.
	\item $h_{d,i}$ is the weight guess $i$ for trace $d$.
	\item $\overline{t_{j}}$ is the mean of all guesses for each trace $d$.
	\item $\overline{h_{i}}$ is the mean of all guesses for a weight $i$.
\end{itemize}

The attacker then uses the absolute value of the PCC to perform the correlation.
A value of $|\rho_{t,h}|$ close to one means that the weight guess $h$ correlates closely with the trace $t$, indicating that weight guess is more likely to be the correct guess for that weight. 
The value with the highest $|\rho|$ is taken as the final guess for that weight.
This process is then repeated for every weight in the trace to generate all the weights for the network.
In CSI NN, the authors are able to reverse engineer networks with a $<1\%$ loss in classification accuracy.
In contrast, when shuffling is applied using \name, the recovered weights yield a network with a much lower classification accuracy.
For one of the networks we evaluate in Section~\ref{s:system-evaluation}, the accuracy using the recovered weights is just 11.7\%. 

\myparagraph{Number of layers.}
Figure~\ref{figure:csi-neuron} shows a single fully connected layer with six neurons.
However, it is not possible to tell this network apart from a network with two layers having three neurons each.
In CSI NN, the authors use the PCC values to also determine the layer boundaries.
The attacker uses a known input to attack all the neurons in the trace.
The neurons belonging to the first hidden layer will correlate strongly with the input (i.e., have high PCC values).
However, as neurons in the second hidden layer do not depend on the input, they show weak correlation. 
Thus, the last neuron which shows a high correlation marks the layer boundary.

The attacker follows an iterative procedure where they target the first hidden layer, determine its size and recover the weights. 
Once this is done, they can calculate the outputs of that layer and feed them to the second hidden layer as inputs and repeat the attack. 
The attacker repeats this process for each layer to reverse engineer the whole network.
We now describe how we reproduce the CSI NN attack as a baseline to evaluate \name.

\subsection{Reproducing the attack}
\label{s:attack-reproduce}

To reproduce the CSI NN attack, we use the ChipWhisperer CW-NANO platform~\cite{chipwhisperer2019}.
The ChipWhisperer is a commonly used platform for side-channel analysis~\cite{Lagasse2019,Golder2019,Mpalane2016}. 
The CW-NANO platform consists of an ARM M0+ CPU as the `target' for side channel attacks, alongside an FPGA for data collection and processing. 

We collect traces of a MLP network consisting of a 32, 10 and 5 neurons in the input, hidden and output layers, respectively. 
We first split this trace into segments of just the weights for each neuron and then use correlation power analysis (CPA) to reverse engineer the weights. 
We empirically determine that 100 traces is sufficient to recover all the weights of the network with 100\% accuracy. 
Our experiments differ from those in CSI NN in two ways: 
We are able to recover the weights with 100\% accuracy with just 100 traces in contrast to CSI NN, which required 
In contrast, CSI NN required several hundred traces and weights were not recovered with 100\% accuracy.
This is due to the following two reasons: 1)~CSI NN targeted floating-point values while we target fixed-point operations.
Low-power IoT devices typically lack floating-point hardware, which makes fixed-point operations a natural choice for running NNs on these devices.
2)~ CSI NN used the EM side channel which is more susceptible to noise compared to the power side channel which we use. 
Despite this, our attack is equivalent to CSI NN since EM is merely a proxy for power.
We also see that the number of traces needed does not scale with the number of weights.
This is because each weight is treated independently, thus having more weights does not affect the `averaging of traces' needed to recover each weight value. 

\subsection{Extending the attack}
\label{s:attack-extending}

While CSI NN targets MLPs and CNNs, we show that this attack also works for other ML algorithms, namely autoencoder (AE) networks and support vector machines (SVMs).
As AE networks use the same layer types as CNNs, the attack applies directly to them. 
For SVMs, we target linear kernels (suitable for low power IoT devices), which use two nested for loops.
The outer loops iterates over all support vectors and the inner loops over all the input dimensions.
The inner loop performs a dot product of the input and a secret weight vector, which an attacker wishes to steal.
Thus, shuffling SVMs is similar to shuffling fully connected layers, which are also implemented using two nested for loops.
Next, we demonstrate for the first time how shuffling performed in software can still be attacked via side-channel information.

\section{Attacking Software Shuffling}
\label{s:software-shuffling}

In this section, we describe how shuffling is implemented in software and how this implementation leaks side channel information. 
Finally, we outline our attack against software shuffling, which can nullify the security benefits of shuffling in software.

\subsection{Shuffling for security}
\label{s:shuffling-background}

\myparagraph{Shuffling for neural networks.}
Software shuffling has been applied to prevent side channel attacks against neural networks~\cite{Brosch2022, Nozaki2021}.
Both papers shuffle the order of neurons per layer as well as the order of weights per neuron.
Algorithm~\ref{a:sw-shuffling} shows a shuffled implementation of a fully-connected layer with $M$ neurons and $N$ weights per neuron.
In the un-shuffled case, the next neuron to run is picked by the loop iterator $i$. 
With shuffling, we need a separate list to store the shuffled order.
Therefore, we make a new list with the values $[0,M)$ in sequence, using the \texttt{CreateList} function (Line~\ref{a:line:SW-create-list}).
The new list is then shuffled and for each loop iteration, we read the next element from the shuffled list and run that neuron (Line~\ref{a:line:SW-random-index}).
This process is repeated each time this layer is run, effectively randomizing the order of operations.
The weights per neuron are also shuffled in a similar way. 
Next, we describe how the shuffled list is created in software.

\begin{algorithm}
	\DontPrintSemicolon
	M\_list = CreateList(M)\; \label{a:line:SW-create-list}	
	M\_shuffled = FisherYatesShuffle(M\_list)\; \label{a:line:SW-shuffle-list}	
	\For{$i = 0;\ i < M;\ i++$}{
		r\_i = M\_shuffled[i]\; \label{a:line:SW-random-index}
		N\_list = CreateList(N)\;	
		N\_shuffled = FisherYatesShuffle(N\_list, N)\;			
		\For{$j = 0;\ j < N;\ j++$}{			
			r\_j = N\_shuffled[j]\;
			sum[r\_i] += input[r\_j] $\times$ weight[r\_i][r\_j]\;
		}
		sum[r\_i] += bias[r\_i]\;			
		output[r\_i] = actFunc(sum[r\_i])\; 
	}
	\caption{Fully connected layer with software shuffling.}
	\label{a:sw-shuffling}
\end{algorithm}

\myparagraph{Fisher-Yates shuffling.}
Prior work uses the Fisher-Yates algorithm (Algorithm~\ref{a:fisher-yates}) for shuffling~\cite{Brosch2022}.
The Fisher-Yates algorithm is widely used to perform shuffling in security-critical applications such as data and image encryption~\cite{Ahmad2014, Tayel2018, Hazra2016, Saeed2014}.
Given a list of $N$ numbers, Algorithm~\ref{a:fisher-yates} generates a random permutation of this list.
Algorithm~\ref{a:fisher-yates} iterates over every item in the list and for each item, picks a second random item and swaps them. 
The $rand()$ function queries a $l$-bit TRNG, which produces a number in the range $[0,2^l)$ (Line~\ref{a:line:FY-rand}).
The TRNG output is scaled to the desired range of $[0,i+1)$ with a modulus operation. 
Finally, the $swap()$ function then swaps both entries (Line~\ref{a:line:FY-swap}).
When the all iterations are complete, the items in the list indicate the random order in which iterations should be run.

\begin{algorithm}
	\DontPrintSemicolon
	\SetKwFunction{FMain}{FisherYatesShuffle}
	\SetKwProg{Fn}{Function}{:}{}
	\Fn{\FMain{$list$, $N$}}{
		\For{$i = N-1;\ i > 0;\ i--$}{
			j = rand() \% (i+1); \label{a:line:FY-rand} \\
			swap(list[i], list[j]);	\label{a:line:FY-swap}		
		}
	}
	\caption{Fisher-Yates algorithm for shuffling.}
	\label{a:fisher-yates}
\end{algorithm}

\myparagraph{Computing modulus.}
We now focus on the modulus operation, which is the source of the side channel leakage. 
In hardware, modulus is computed as the remainder of a division operation~\cite{ARMDivide}.
Ultra-low power CPUs, such as the ARM M0+ that we use in our evaluation, do not have a hardware divider~\cite{ARMDivide}.
They instead implement division in software, using shifts and subtracts. 

\begin{algorithm}
	\DontPrintSemicolon
	\SetKwFunction{FMain}{division}
	\SetKwProg{Fn}{Function}{:}{\KwRet}
	\Fn{\FMain{$a$, $b$}}{
		\uIf{$b$ == $0$}{ 						\label{a:line:div-check-zero}
			$divideByZeroException()$ \;
		}
		\Else{		 
			$i = 1$, $q = 0$\;					
			\While{$b[31]\neq 0$}{				\label{a:line:div-count-bits}				 
				$b = b << 1$ ; $i = i << 1$\;
			}
			\While{$i>0$}{						\label{a:line:div-calculate}
				$q = q << 1$ \;
				\If{$a \geq b$}{
					$a = a-b$ \;
					$q = q+1$ \;
				}			
				$b = b >> 1$ ; $i = i >> 1$\;
			}			
		}		
	}
	\Return{$q$, $a$} \;
	\caption{Pseudocode for software division.}
	\label{a:sw-divide}
\end{algorithm}

\myparagraph{Software division.}
For the M0+ CPU, ARM GCC (i.e., \textit{arm-none-eabi-gcc}) uses the \verb|__aeabi_udivmod| function for division and modulus.
Algorithm~\ref{a:sw-divide} shows pseudo-code for this function.
The $division$ function computes $a\div b$ and returns the quotient $q$ and remainder (i.e., the modulus) $r$.
The first \textit{While} loop counts the number of steps division will take, by shifting $b$ 1-bit to the left until bit 31 is $1$.
The number of shifts required is stored in $i$, which then determines how many times the second \textit{While} loop runs.
The second \textit{While} loop performs division by implementing a \textit{restoring division} algorithm.
Both the time taken and the power trace vary based on the dividend $a$ and divisor $b$.

\subsection{Analyzing software division}
\label{s:software-analyzing}

\begin{figure*}[t!]
	\centering
	\begin{subfigure}[t]{0.39\textwidth}
		\centering
		\includegraphics[width=0.99\columnwidth]{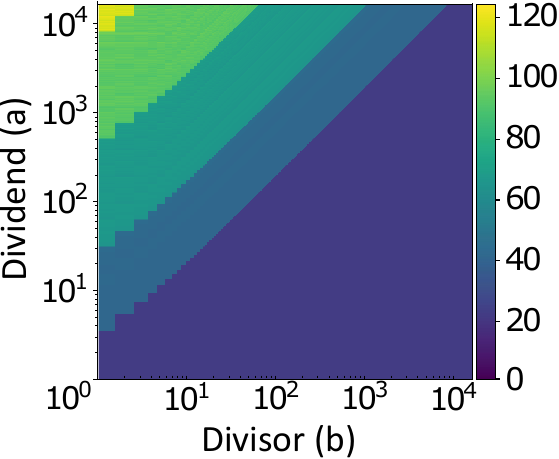}
		\caption{Heat-map of cycle times for software division, using a log-log scale.}
		\label{figure:swdiv-heatmap}
	\end{subfigure}	
	\quad
	\begin{subfigure}[t]{0.58\textwidth}
		\centering
		\includegraphics[width=0.99\columnwidth]{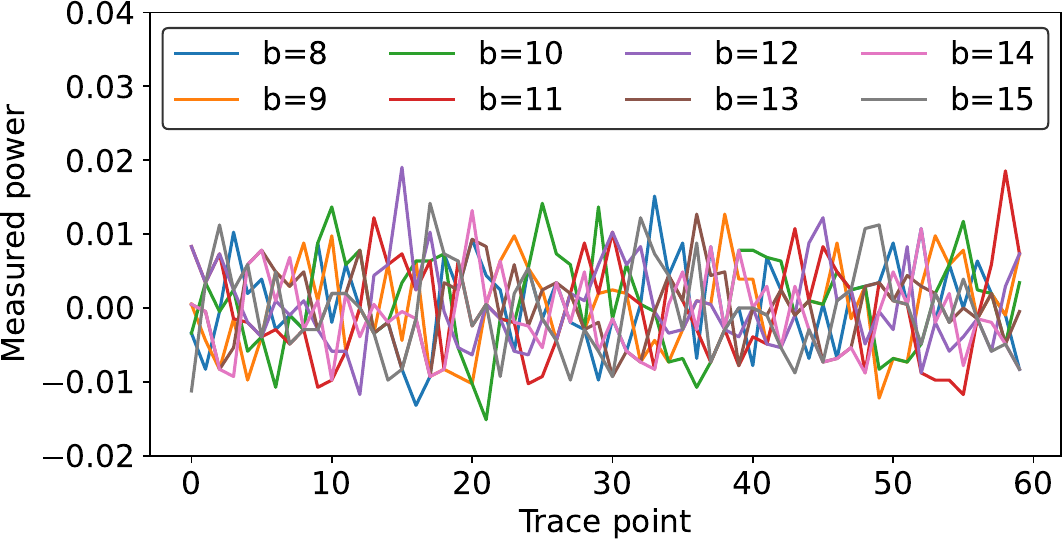}
		\caption{Mean subtracted power traces of $100\div b$ for $b\in[8, 15]$.}
		\label{figure:swdiv-power}
	\end{subfigure}	
	
	\begin{subfigure}[t]{0.39\textwidth}
		\centering
		\includegraphics[width=0.99\columnwidth]{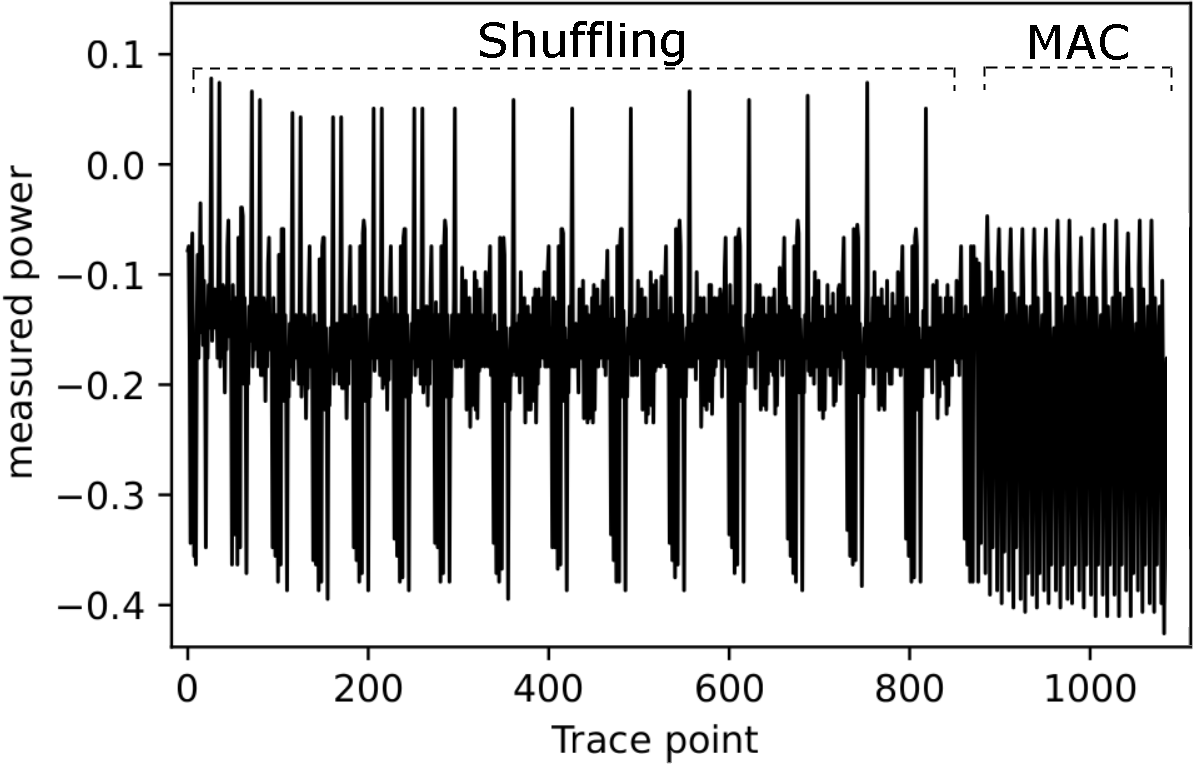}
		\caption{Trace showing software shuffling and multiply-accumulate operations for a neuron with 16 weights.}
		\label{figure:swshuffle-trace}
	\end{subfigure}
	\hfill
	\begin{subfigure}[t]{0.56\textwidth}
		\includegraphics[width=0.99\columnwidth]{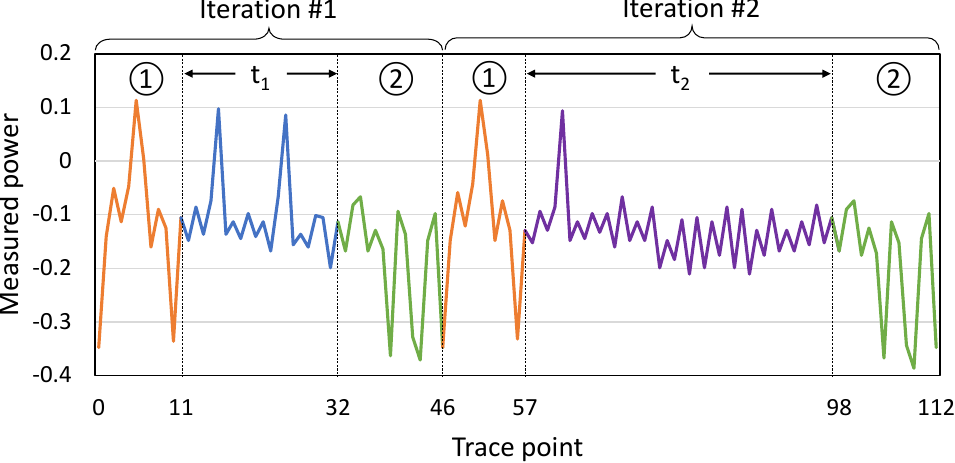}
		\caption{Splitting a trace of two loops of software shuffling.}
		\label{figure:swshuffle-splitting}		
	\end{subfigure}
	\caption{Analysis for side-channel attack against software division.}	
\end{figure*}

\myparagraph{Latency variation.}
We begin by profiling the number of cycles taken by software division.
We once again use the CW-NANO platform we described in Section~\ref{s:attack-reproduce}.
We measure latency using a C program, compiled using the ARM GNU compiler v9.2.1 with \textit{-O3} optimizations.
Figure~\ref{figure:swdiv-heatmap} shows the heat-map of cycles of $a\div b$ for $a,b \in [1, 16384)$.\footnote{We use 16,384 as that is the maximum number of iterations our implementation supports (Section~\ref{s:system-evaluation}). However, our attack scales to all values of $a$ and $b$.}
We see a significant variation in latency when $a>b$ (top left of Figure~\ref{figure:swdiv-heatmap}).
As Algorithm~\ref{a:sw-divide} performs $a-b$ during each iteration, the bigger then value of $a$ compared to $b$, the more iterations are needed. 
In contrast, the latency is similar for all cases where $a<b$ (bottom right of Figure~\ref{figure:swdiv-heatmap}).
This is because in these cases, Algorithm~\ref{a:sw-divide} only runs a single iteration.
Since many input values have the same latency, we also analyze the variation in \textit{power} when performing division, to uniquely identify $a$ and $b$. 

\myparagraph{Power variation.}
Figure~\ref{figure:swdiv-power} shows mean subtracted power traces for $100\div b$ for $b\in[8, 15]$.
We first take the average of all the power traces (to remove the power contribution of other system components) and then plot each trace minus this average.
While dividing $100$ by each of these $b$ values has the same latency, we see that the power traces differ based on the value of $b$, allowing us to tell them apart.
While we only show a small range of values for clarity, we see this behaviour for the entire range of inputs we study.
Together, we use the input-dependent variation in the latency and power of software division as the basis of our attack.

We could gather and store traces of the system computing $a\div b$ for all possible values of $(a,b)$. 
Then, we would compare each stored trace against each trace we collect from the system. 
The stored trace which exactly matches the collected trace would give us the values of $a$ and $b$.
However, this approach suffers from two drawbacks: 
1)~as $a$ and $b$ can each take $2^{N}$ values, the number of comparisons required grows exponentially with $N$.
2)~As Figure~\ref{figure:swdiv-heatmap} shows, many values of $a$ and $b$ have the same latency. 
Thus, the variations in the power traces between these values is small, making it difficult to uniquely identifying $a$ and $b$ from a single trace this way. 
We now describe two techniques to narrow the search space and make this identification tractable.

\myparagraph{1. Making efficient comparisons.}
From our earlier profiling of software division, we have a minimum ($t_{min}$) and maximum ($t_{max}$) time that division can take.
Instead of finding out where division ends, we find where in the collected trace \ding{193} begins.
As \ding{193} is the same for every trace, this comparison is much more efficient.
For the first division operation in Figure~\ref{figure:swshuffle-splitting}, we compute the difference between \ding{193} and the collected trace starting $11+t_{min}$ until $11+t_{max}$.
The value of $t$ (i.e., $t_{div}$) where the difference is $0$ gives us the latency of division. 
Once we know $t_{div}$, we only need to compare against traces which take that number of cycles. 
However, as Figure~\ref{figure:swdiv-heatmap} shows, many values can have the same division latency.
Our second optimization further shrinks the search space.

\myparagraph{2. Sequential values.}
In Algorithm~\ref{a:fisher-yates}, the inputs to the division operation are $rand()$ and $i+1$.
We cannot know $rand()$ as it is the output of a TRNG. 
However, as $i$ goes from $N$ to $1$, we know the value of $i+1$ during each iteration.\footnote{Some implementations of Fisher Yates access items from index $0$ to $N-1$. Our approach still applies as elements are accessed sequentially.}
We can learn $N$ by analyzing the traces shown in Figure~\ref{figure:csi-weight} as software shuffling does not obscure the number of items being shuffled.
We now know the divisor ($b$, which is $i+1$) during each division operation.
We only need to compare the trace against the stored traces where the divisor is $i+1$, which further reduces the number of comparisons needed.

\myparagraph{Training a classifier.}
With the first two optimizations having reduced the number of comparisons needed, we train decision tree classifiers to predict $a$, given $b$ and $t_{div}$.
We train our classifiers using SciPy version 1.9.0, using the \textit{gini} criterion.
We train a separate classifier for each value of $t_{div}$ and $b$.
By narrowing the range of values that each classifier must predict, we obtain smaller and more accurate classifiers. 
This allows the classifier to predict the value of $a$ with 100\% accuracy.

Note that we get $a$ and $b$ from \textbf{a single trace}.
CSI NN requires multiple traces because the multiplication operation being targeted is a single-cycle operation.
However, as software division takes many cycles, our classifier has many data points it can use, which allows our attack to work with a single trace.

\subsection{Putting it all together}

The target of our attack is the modulus operation (Line~\ref{a:line:FY-rand} in Algorithm~\ref{a:fisher-yates}).
We wish to learn the value of $j$ so we know the inputs to the $swap()$ function (Line~\ref{a:line:FY-swap}).
Using our attack, we find the output of $rand()$ and we also know the value of $i$.
Knowing these values lets us determine the value of $j$ for every iteration. 
We then collect multiple traces as outlined in Section~\ref{s:attack} and then rearrange each trace based on the swapped indices.
With the rearranged traces, we can carry out the power side-channel attack as before.
With our attack, software shuffling offers \textbf{no security improvement} over the baseline. 
We therefore conclude from our attack that we require novel hardware which does not leak side channel information for shuffling.

\section{Securing model weights with \name}
\label{s:solution}

In this section, we describe \name, hardware for efficient shuffling. 
We begin by outlining the main challenges associated with designing hardware for shuffling.
We then provide an overview of our hardware, followed by a description of how software interfaces with our hardware.

\subsection{Design challenges}
\label{s:solution-challenges}

\myparagraph{Avoiding the memory bus.}
We want to avoid the memory bus as it is the main source of information leakage.
Therefore, we add \name as a functional unit directly within the CPU.
This also reduces the latency of our approach.

\myparagraph{Reducing latency.}
Similar to software shuffling, we could also produce a shuffled list ahead of each loop.
However, storing a list of arbitrary size $N$ in hardware is challenging.  
We cannot store this list in memory as that would require using the memory bus and subject to leaking information.
Alternatively, we could use a dedicated on-chip storage but sizing this to accommodate the large dimensions of neural networks would add considerable overhead. 
As we show in Section~\ref{s:system-evaluation}, our implementation supports $16,384$ iterations for four loops.
We use CACTI 7~\cite{cacti} to determine that storing this many iterations would add $61$\% area overhead to an ARM M0+ SoC~\cite{Myers2015}.
Instead, we produce random iterations while the layer is running and store the next iteration value in a CPU-accessible register. 
The CPU reads from this register in a single cycle, thereby minimizing latency and storage overhead. 

\myparagraph{Avoiding the modulus operation.}
We must convert the TRNG output from a value in the range $[0, 2^l)$ to the range $[0,N)$, using a modulus operation.
As we showed in Section~\ref{s:software-shuffling}, modulus (implemented as division) is susceptible to side-channel attacks.
Therefore, we need a way to randomize the order of iterations without using a modulus operation.
We now describe \name, which addresses these challenges without incurring significant overheads. 

\begin{figure}
	\centering
	\includegraphics[width=0.99\columnwidth]{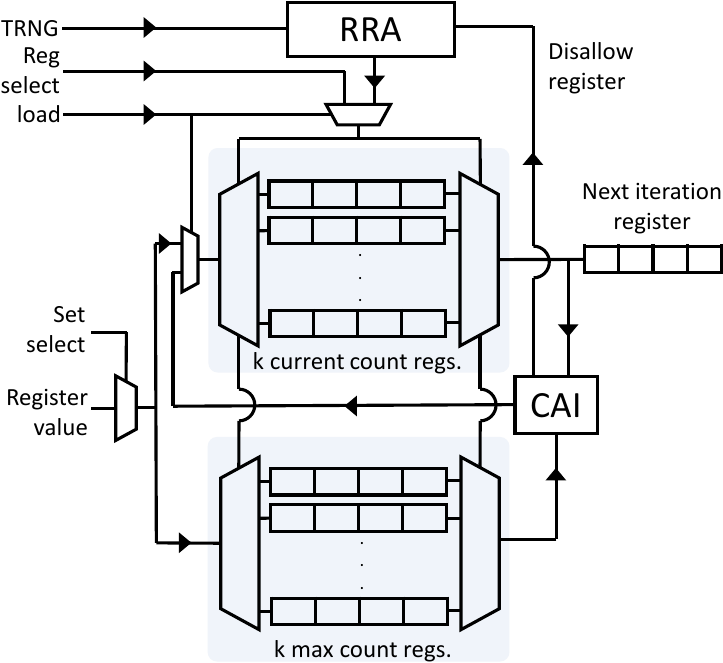}
	\caption{Hardware for counter-based shuffling.}
	\label{figure:hardware-one-bank}
\end{figure}

\begin{algorithm}
	\DontPrintSemicolon
	\colorbox{lightgray}{load\_bank(BANK0,M)}\;
	\colorbox{lightgray}{load\_bank(BANK1,N)}\;
	\For{$i = 0;\ i < M;\ i++$}{
		\colorbox{lightgray}{r\_i = get\_next\_iteration(BANK0)}\;
		\For{$j = 0;\ j < N;\ j++$}{
			\colorbox{lightgray}{r\_j = get\_next\_iteration(BANK1)}\;
			sum[r\_i] += input[r\_j] * weight[r\_i][r\_j]\;
		}
		sum[r\_i] += bias[r\_i]\;			
		output[r\_i] = actFunc(sum[r\_i])\; \label{a:line:act-func}
	}
	\caption{Fully connected layer with \name functions added.}
	\label{a:shuffled-mlp}
\end{algorithm}		

\subsection{High level overview }
\label{s:solution-overview}

To generate iterations in the range $[0,N)$ in a random order, without repetitions, we first split the total number of iterations into $k$ `bins'. 
Each bin then represents a subset of the total number of iterations that must be run.
For example, for a single neuron with ten weights, we split them into two bins: bin 0 for iterations 0-4 and bin 1 for iterations 5-9.\footnote{The number of bins is configurable at design time. We use a design with 2 bins in this example for clarity.} 
To start with, each bin is set to its minimum value (i.e., 0 and 5).
To pick an iteration to run, we pick one of the two bins and the value in that bin is output. 
Next, the value in that bin is incremented, ensuring a unique output each time. 
The process repeats ten times to output ten total iterations, with one of the two bins picked randomly each time.
In essence, \name converts the problem of selecting an iteration in the range $[0,N)$, to picking from a much smaller number of bins.
By restricting the number of bins to always be a power of 2, \textit{we can directly use the output of the TRNG without requiring a modulus operation.} 
Next, we quantify the total number of possible permutations when using \name.

\myparagraph{Mathematical formulation.}
When $N$ is a multiple of $k$, all bins will have the same number of iterations.
But when $N$ is not a multiple of $k$, there will be one bin with fewer iterations. 
In this case, the first $k-1$ bins will each have $a$ iterations where $a = \left \lceil N/k \right \rceil$, while the last bin will have $b$ iterations, where $b = N-(k-1)a$.
If $N$ is a multiple of $k$ however, $a = N/k$ and $b = 0$. 
Therefore, the total number of permutations, $P$ is:
\begin{equation}
	P=
	\begin{cases}
		N!/[(a!)^{k}], & \text{if N is a multiple of k} \\
		N!/[(a!)^{k-1}\times b!], &\text{otherwise}
	\end{cases}
	\label{e:permutations}
\end{equation}

For the case with a single register ($k = 1$), we have $a=N$ (i.e., $N$ iterations all in one bin) and $b=0$. 
This gives us just $1$ possible order, which is the same as the baseline case.
Using $N$ registers per set (i.e., $k = N$), with $a=1$ (i.e., 1 bin per iteration) and $b=0$, gives us the maximum possible $N!$ permutations.
Using Equation~\ref{e:permutations} for our example above (N=$10$ and k=$2$), we get $P=252$ possible orderings.
For larger sizes of $N$ and $k$, $P$ quickly grows into the millions, which effectively randomizes the sequence. 
In Section~\ref{s:evaluation-quantification}, we show how such huge values of $P$ make the attack take intractable lengths of time.
We now describe our hardware implementation of this `bins' to track iterations.

\subsection{Hardware overview}
\label{s:solution-hardware}

Figure~\ref{figure:hardware-one-bank} shows an overview of \name. 
We use a set of $k$ registers to keep track of the value of each bin.
In our example above, we would use two \textit{current count} registers, to track the current value of each bin. 
These two current count registers are initially loaded with the values 0 and 5, respectively. 
We use a TRNG to pick a current count register to output the next iteration.
As registers are picked, their values are incremented each time.
However, once a current count register reaches its maximum value (i.e., 4 or 9 in our example), we must disallow it from being run again.
To keep track of the maximum values for each current count register, we use another set of \textit{max count} registers.

As current count registers begin to saturate, the output from the TRNG will pick disallowed registers.
To quickly pick another valid register to run, we employ a combinational Round Robin Arbiter (RRA). 
The RRA keeps track of all current count registers, using a single bit set to `1' per register, indicating that this current count register still has iterations that can be run. 
The output of the TRNG is fed to the RRA to pick a current count register.
If the corresponding RRA bit is `1', that register's value is output. 
Next, we compare the value in that current count register with its corresponding max count register. 
If the max value has not been reached, the current count register is incremented.
This is performed using the `compare and increment' (CAI) block in Figure~\ref{figure:hardware-one-bank}.
However, if a register has reached its maximum value, the CAI block sets the bit corresponding to that register in the RRA to `0', via the `disallow register' signal.
If a disallowed register is later picked, the RRA outputs the closest allowable register to be run instead. 
The RRA is purely combinational and therefore returns a valid register in a single cycle each time.\footnote{This avoids repeatedly querying the TRNG for a valid register which would be time consuming.}
The number of registers per set is a parameter that can be configured at design time. 
We explore both the frequency of our design and the number of register per set we use in Section~\ref{s:evaluation}. 
The larger the number of registers, the more security our design provides but at the cost of increased area.
To balance security and added area, we opt for $16$ registers per set. 

\myparagraph{Hardware banks.}
The hardware shown in Figure~\ref{figure:hardware-one-bank} generates random iterations for a single loop.
However, neural network layers are implemented as a series of nested loops.
We therefore use one copy of the hardware in Figure~\ref{figure:hardware-one-bank} per loop that we wish to randomize. 
Each loop is associated with one bank and we use multiplexors to pick which bank to use for each loop.
We opt for a design which uses four banks, to balance security vs.~area and latency overhead. 
We use two banks for fully connected layers.
For convolutional layers, we opt to four out of the six loops.
Therefore, we loop over input channels, the output channels, input rows and input columns.
Lastly, for max pooling layers, we use three banks to shuffle rows, columns and channels.
Shuffling of these loops is achieved by means of additions we make to the code (shown using highlighted boxes in Algorithm~\ref{a:shuffled-mlp}). 
Next, we describe the purpose of these code additions.

\subsection{System interface}
\label{s:solution-interface}

In this section, we show how \name is controlled via software and the necessary extensions to support this.

\myparagraph{Code annotations.}
We program \name via two functions: $load\_bank$ and $get\_next\_iteration$.
Algorithm~\ref{a:shuffled-mlp} shows the code for a fully connected layer with our changes highlighted. 
The $load\_bank$ function (lines 1 and 2) loads the registers in a specified bank, before running the loops. 
The $get\_next\_iteration$ function (lines 4 and 6) queries the hardware for the next iteration from a given bank. 
The values returned from the $get\_next\_iteration$ are stored in $r\_i$ and $r\_j$ and then used in the loops instead of the original loop iterators $i$ and $j$. 
The $load\_bank$ and $get\_next\_iteration$ functions are defined in a library that we provide. 
Our library implements these functions using custom ISA instructions, which we describe next.

\begin{figure}[t!]
	\centering
	\includegraphics[width=0.99\columnwidth]{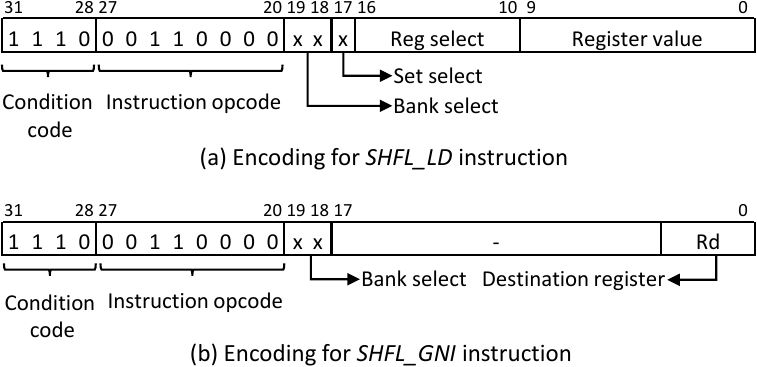}
	\caption{Custom ISA instructions for shuffling hardware.}
	\label{figure:isa-codes}
\end{figure}

\myparagraph{ISA extensions.}
We add additional CPU instructions to interface with \name (Figure~\ref{figure:isa-codes}).
The first instruction, \textit{SHFL\_LD}, loads initial values to the current count and max count registers before each layer.
The bits of the \textit{SHFL\_LD} instruction are:
\begin{itemize}[leftmargin=*, noitemsep]
	\item {[31:28]} represent condition codes that the instruction must check before execution.
	We set these bits to `1110', as per the ARM Technical Reference Manual~\cite{ARM_manual}.	
	\item {[27:20]} shows an unused opcode in the baseline ARM ISA which we use for our instructions.
	\item {[19:18]} select the bank we want to access.
	\item {[17]} select the set (i.e., current/max count registers).
	\item {[16:10]} specify the register within the set. 
	\item {[9:0]} are the value to be loaded into the selected register.
\end{itemize}

The second instruction, \textit{SHFL\_GNI}, returns the next iteration from one of the banks.
This instruction format is:
\begin{itemize}[leftmargin=*, noitemsep]
	\item {[31:18]} are identical to the \textit{SHFL\_LD} instruction.
	\item {[17:4]} are unused in this instruction.
	\item {[3:0]} specifies a CPU register for the result.
\end{itemize}

The $SHFL\_LD$ instruction uses 10 bits for the register value, which allows each register to count up to 1024 iterations.
We use 7 bits for the register select, which allows for designs with up to 128 registers per set.
This instruction encoding therefore supports loops with up to 131,072 iterations.
As this is much larger than networks run on an IoT device, this encoding does not limit the size of networks that our technique can support.\footnote{For example, the largest layer we run in Section~\ref{s:solution-latency} is an order of magnitude smaller than the maximum iterations supported by our encoding.} 
Our technique does not impose any restriction on the number of layers nor the total number of weights a network can have. 

Our library contains definitions for the $load\_bank$ and $get\_next\_iteration$ function calls.\footnote{Our library can be integrated within other NN libraries such as STMicro's STM32Cube.AI~\cite{STM32CubeAI}.}
The $load\_bank$ function calculates and loads (using \textit{SHFL\_LD}) the current count registers and max count registers.
The $load\_bank$ function is only called once per bank, before each layer.
The overhead of $load\_bank$ scales with the number of registers but is not affected by the size of the layers.
Thus, layers of any size require the same number of instructions for the loading operation, which amortizes the overhead of $load\_bank$.

The $get\_next\_iteration$ function performs a single register read and therefore adds just one \textit{SHFL\_GNI} instruction to the program binary. 
However, as the \textit{SHFL\_GNI} instruction is called for every single loop iteration, it is critical that we minimize the cycle count of that instruction, to reduce the overall latency impact.
Next, we explain how \name achieves this goal of minimizing the latency of the \textit{SHFL\_GNI} instruction.

\subsection{Hardware latency}
\label{s:solution-latency}
We design \name to provide the next iteration number to the CPU with a one-cycle latency. 
To do this, we take advantage of the time between subsequent calls to the $get\_next\_iteration$.
In Algorithm~\ref{a:shuffled-mlp}, we first load BANK0 (line 1). 
As soon as BANK0 is loaded, the hardware begins selecting the next iteration for that bank.
In the meantime, the CPU is loading values for BANK1 (line 2).
Thus, we have several cycles to pick the next iteration for BANK0 before it is queried by the CPU.

Similarly, there are seven cycles between subsequent calls to the $get\_next\_iteration$ function, even in the inner for loop (line 5). 
This is because CPU must do several operations (i.e., calculating the index of the next weight, loading that weight and associated input, performing a multiplication and addition) before requiring the next iteration number.
This allows \name to select the next iteration in time for the next request from the CPU. 
\name takes a total of three cycles to generate the next iteration, which gives us several cycles of buffer before the next iteration is queried.

Once calculated, the next iteration value is stored in the `next iteration' register in each bank until it is read by the CPU. 
The CPU can then read from this register using the \textit{SHFL\_GNI} instruction in a single cycle. 
As soon as this register is read, the hardware begins selecting the next iteration to run for that bank.
This allows us to minimize the latency of the $get\_next\_iteration$ to a single cycle.

\section{Evaluation}
\label{s:evaluation}

In this section, we evaluate the effectiveness of \name in preventing side-channel attacks. 
We begin by showing that \name masks the side channel and secures neural networks against attackers.
Next, we show how \name greatly increases the time needed to collect enough traces to carry out the attack.
We then report the runtime overhead of \name on several representative neural networks and compare against the overhead of software shuffling.
Finally, we quantify the area and latency overhead of \name and explain how \name does not leak side channel information.

\begin{figure*}
	\centering
	\begin{minipage}{0.99\textwidth}
		\centering
		\includegraphics[width=0.99\textwidth]{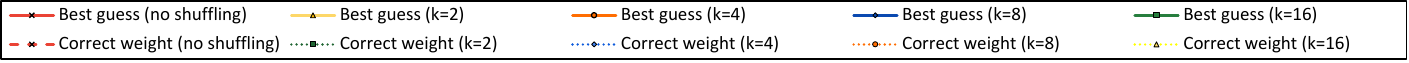}
	\end{minipage}

	\begin{minipage}{0.99\textwidth}
		\centering
		\begin{subfigure}[t]{0.48\textwidth}
			\includegraphics[width=0.99\textwidth]{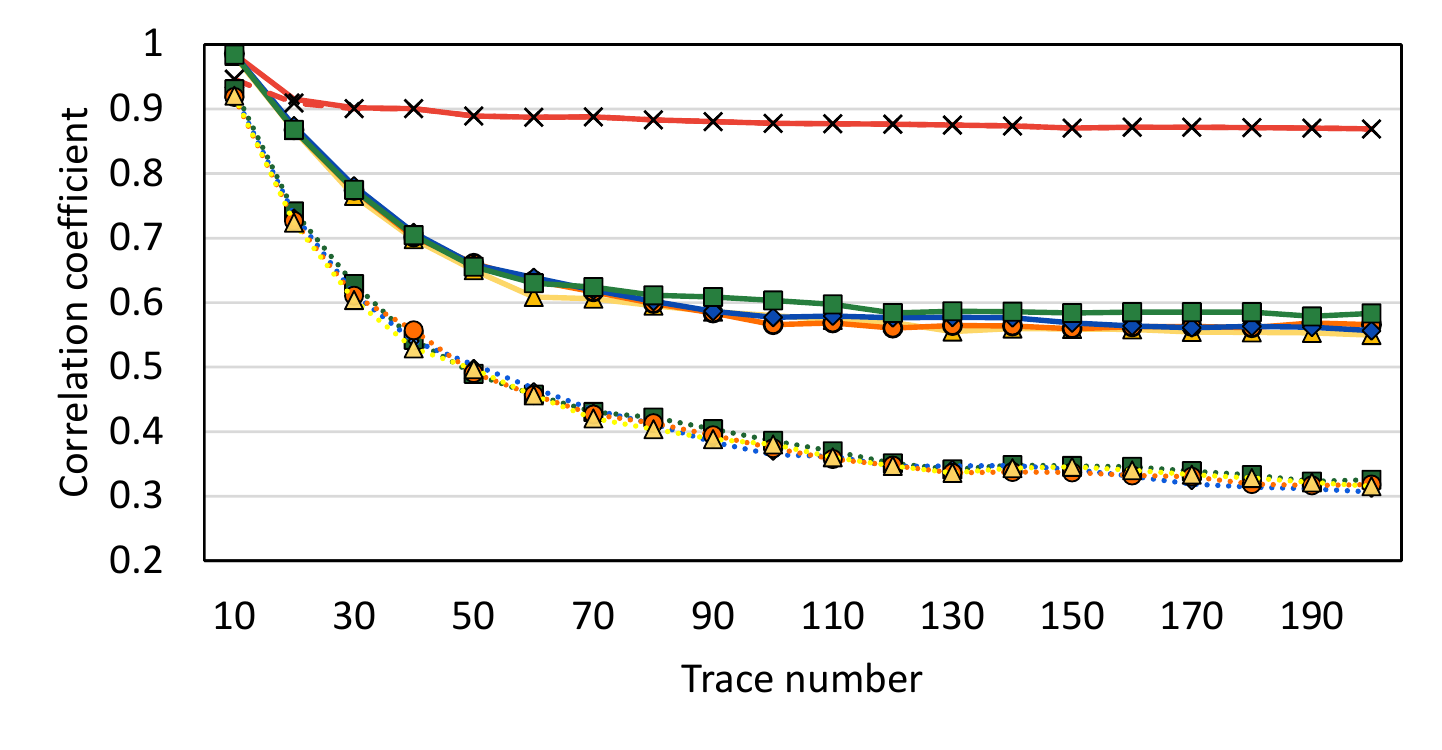}
			\caption{Average correlation coefficient ($\rho$)}
			\label{figure:cpa-values}
		\end{subfigure}
		\begin{subfigure}[t]{0.48\textwidth}
			\includegraphics[width=0.99\textwidth]{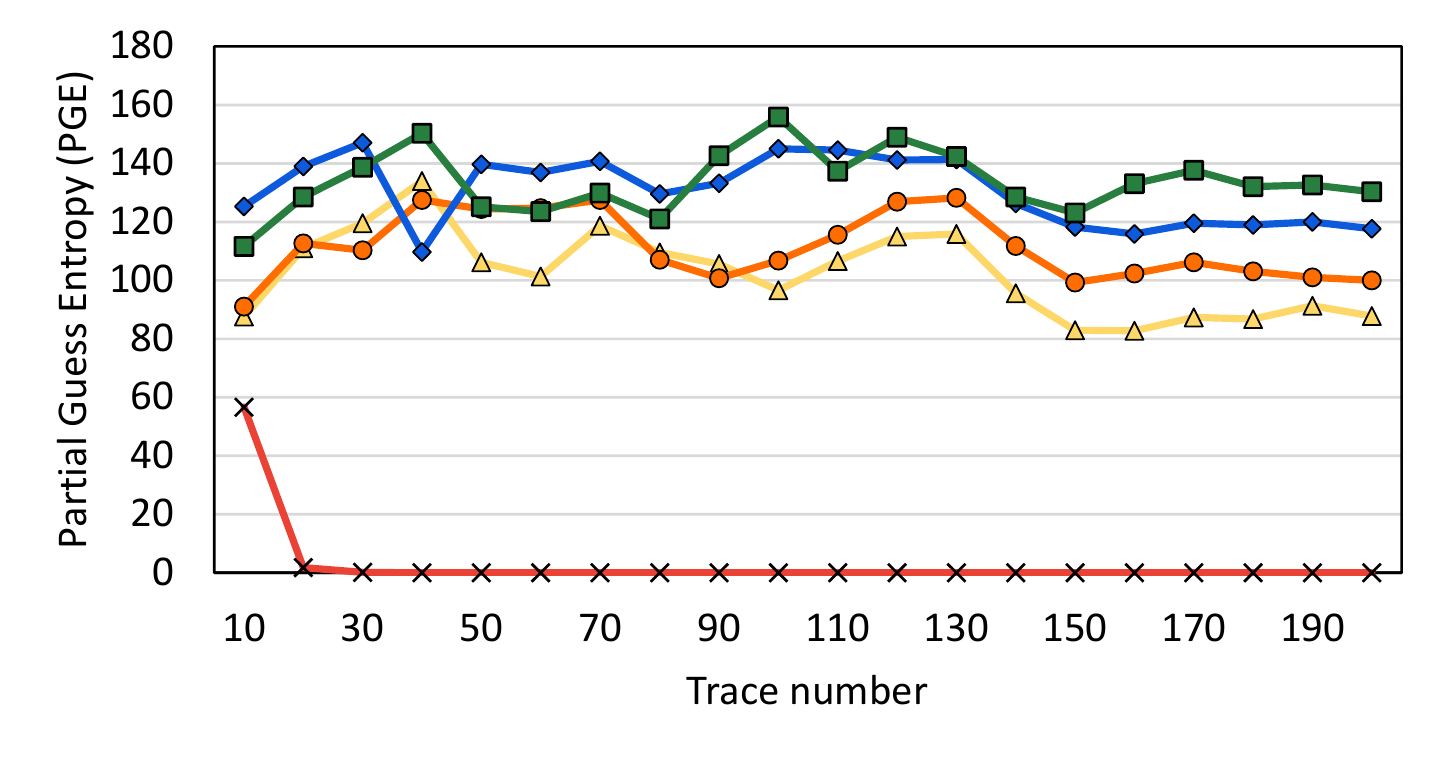}
			\caption{Average Partial Guess Entropy (PGE)}
			\label{figure:pge-values}
		\end{subfigure}
	\caption{Effect of different degrees of shuffling for 16 weights.}		
	\end{minipage}
\end{figure*}

\subsection{Efficacy of hardware shuffling}
\label{ss:efficacy}

We first show how shuffling impacts the effectiveness of the power side-channel attack. 
Since our target CPU does not have shuffling hardware, we calculate a shuffled order of weight accesses for each run and load this into our CPU prior to trace collection. 
Thus the traces we collect have the weights accessed in a new shuffled order during each run.
For robustness, we collect 200 traces for each run of the network, exceeding the 100 traces we need for the baseline attack.\footnote{Our experiments show that even increasing the number to 1000 traces does not change our results.}
This ensures that merely increasing the number of traces required does not allow an attacker to circumvent our solution.

\myparagraph{Effect on $\rho$.}
First, we see how shuffling affects the Pearson correlation coefficient ($\rho$), which is the metric used by CPA to determine the most likely value of each weight. 
We demonstrate using a single run with 16 weights, where we vary the amount of shuffling from $k=1$ (no shuffling) to $k=16$ (full shuffling). 
Recall that $k$ is the number of registers per set in our design. 
Figure~\ref{figure:cpa-values} shows $\rho$ (y-axis) as we analyze more traces (x-axis) 
For each $k$ value, we show the average $\rho$ of all 16 weights for two cases: 
\begin{enumerate}[leftmargin=*]
	\item \textit{Continuous line}: the weight with the highest $\rho$ value ($\rho_{max}$), which is the weight guessed by the CPA attack 
	\item \textit{Dashed line}: the $\rho$ value of the correct weight ($\rho_{correct}$). 
\end{enumerate}

As we use the Pearson Correlation Coefficient, all weights starts with a value of $1$.
But as we add more traces, we expect $\rho_{correct}$ to stay high while the $\rho$ values for incorrect weights to settle to significantly lower values. 
This is precisely what we see in the no shuffling case, as the continuous and dashed lines overlap.
This means that $\rho_{max} = \rho_{correct}$ and that the attack can identify the correct weight values in just 30 traces. 
However, for cases with shuffling, as we analyze more traces, the attack always guesses an incorrect weight as the best guess (i.e., $\rho_{max}$).
The correct weight ($\rho_{correct}$) is consistently lower, giving the attacker no means of identifying this as the correct weight. 

Despite this, the incorrect guesses could still be numerically close to the correct weights. 
To study this, we use the weights obtained with shuffling for one of the networks we study in Section~\ref{s:system-evaluation}, namely \textit{mnist-mlp}.
With shuffling, the network achieves a classification accuracy of just 11.7\%, compared to the original accuracy of 92.9\%.
Thus, the weights recovered with shuffling do not provide any useful information to the attacker to steal the network.

\myparagraph{Partial Guess Entropy (PGE).}
In addition to looking at $\rho$, we also look at how far away the guessed weight is from the correct weight.
Recall that the CPA attack generates a list of possible weight guesses, ranked by $\rho$.
PGE~\cite{Dofe2016} is the position of the correct weight in this list of guessed weights; A PGE of $0$ means that the attack correctly guessed the weight.
Figure~\ref{figure:pge-values} shows the average PGE values for 16 weights as we vary $k$. 
As expected, with no shuffling, PGE reaches $0$ with just 30 traces analyzed.
However with shuffling, PGE values remain high and do not move closer to $0$, even with more traces. 
This shows that analyzing additional traces does not diminish the effectiveness of our technique.
We also see that increasing $k$ leads to an increase in the average PGE value.
Past 150 traces, we see that the PGE values stabilize in order of $k$, with $k=2$ and $k=16$ having the lowest and highest average PGE values, respectively.
This shows that increasing $k$ leads to an increase in the security offered by our design.
So far, we have used small values of $N$ and $k$ for clarity.
We now quantify the impact of larger values of $N$ and $k$ on the time needed for a successful attack.

\begin{table}[t]
	\centering
	\small
	\renewcommand{\arraystretch}{1.2}
		\caption{Time needed (in years) to collect and process enough traces to carry out the attack for a single dimension, for different values of $N$ (columns) and $k$ (rows).}				
		\begin{tabular}{|c|c|c|c|}
			\hline
			\backslashbox{k}{N} & 32 & 64 & 128 \\ \hline
			2 & 1.91E-02 & 5.81E+07 & 7.59E+26 \\ \hline
			4 & 3.16E+06 & 2.10E+25 & 2.55E+63 \\ \hline
			8 & 7.58E+13 & 5.76E+41 & 3.33E+98 \\ \hline
			16 & 1.27E+20 & 3.32E+56 & 2.51E+131 \\ \hline
			32 & 8.34E+24 & 9.37E+68 & 8.33E+160 \\ \hline
			64 & - & 4.02E+78 & 6.63E+185 \\ \hline
			128 & - & - & 1.22E+205 \\ \hline
		\end{tabular}
	\label{t:time}
\end{table}

\subsection{Effect on time needed for attack}
\label{s:evaluation-quantification}

The increase in number of total permutations is effective as a security measure since it dramatically increases the time needed by the attacker to collect and process enough traces to find the correct weights. 
Table~\ref{t:time} shows the time it would take \emph{(in years)} for an attacker to gather enough traces and process them to recover the weights.
We assume an attacker who can gather and process 1000 traces a second, which is similar to the speed of our setup.
While the time is relatively short for small values (e.g., $\sim7$ days for $N=32, k=2$), this rapidly grows into decades and then centuries for larger values of $N$ and $k$. 
As the benchmarks we evaluate below contain thousands of weights, the time needed to carry out the attack would be thousands of years to reverse engineer even a single layer.
We further note that the analysis above is for securing one single dimension, such as the weights of a single neuron. 
Thus, the time needed to reverse engineer a whole network would be cumulative, making it totally untenable to carry out the attack in a reasonable amount of time.
This tremendous increase in time needed for the attack is the cornerstone of the security offered by \name. 

\subsection{System evaluation}
\label{s:system-evaluation}

\begin{table*}[t!]
	\centering
	\small
	\renewcommand{\arraystretch}{1.2}
		\caption{List of networks evaluated, showing the architecture, and overheads (\texttt{C}--Convolutional and \texttt{F}--Fully connected layers).}
		\begin{tabular}{|cc|cc|}
			\hline
			\multicolumn{1}{|c|}{\multirow{2}{*}{Network}} & \multirow{2}{*}{Architecture} & \multicolumn{2}{c|}{Overhead} \\ \cline{3-4} 
			\multicolumn{1}{|c|}{} &  & \multicolumn{1}{l|}{Software} & \multicolumn{1}{l|}{Hardware} \\ \hline
			\multicolumn{1}{|c|}{mnist-mlp} & F(768$\times$128), F(128$\times$10) & \multicolumn{1}{c|}{75.82\%} & 1.15\% \\ \hline
			\multicolumn{1}{|c|}{kws-mlp} & F(250$\times$144),  F(144$\times$144), F(144$\times$10) & \multicolumn{1}{c|}{59.10\%} & 1.12\% \\ \hline
			\multicolumn{1}{|c|}{mnist-cnn} & C(3$\times$3$\times$1$\times$6), C(3$\times$3$\times$6$\times$6), F(150$\times$20), F(20$\times$10) & \multicolumn{1}{c|}{39.90\%} & 0.19\% \\ \hline
			\multicolumn{1}{|c|}{har-cnn} & C(2$\times$2$\times$1$\times$128), F(5632$\times$128), F(128$\times$128), F(128$\times$6) & \multicolumn{1}{c|}{271.36\%} & 0.21\% \\ \hline
			\multicolumn{1}{|c|}{gesture-cnn} & C(5$\times$5$\times$1$\times$32), C(3$\times$3$\times$32$\times$64), C(3$\times$3$\times$64$\times$64), F(5760$\times$128), F(128$\times$10) & \multicolumn{1}{c|}{100.06\%} & 0.14\% \\ \hline
			\multicolumn{1}{|c|}{ecg-ae} & F(128$\times$1024), F(1024$\times$1024), F(1024$\times$140) & \multicolumn{1}{c|}{75.06\%} & 1.17\% \\ \hline
			\multicolumn{1}{|c|}{seizure-svm} & F(2854$\times$179) & \multicolumn{1}{c|}{86.74\%} & 0.58\% \\ \hline
			\multicolumn{2}{|r|}{Average} & \multicolumn{1}{r|}{101.15\%} & 0.56\% \\ \hline
		\end{tabular}
	\label{t:networks}
\end{table*}

In this section, we evaluate the overhead of shuffling using the networks listed in Table~\ref{t:networks}. 
For each network, we list the layers, the activation function used per layer and the size of each layer.
For FC layers, the size is given as input channels $\times$ output channels, while for CONV layers it is given as kernel width $\times$ kernel height $\times$ input channels $\times$ output channels.
Also, each CONV layer is followed by a 2$\times$2 max pooling layer. 
The networks are written in C and compiled using ARM GCC 2019.4 compiler, with optimization set to -O3. 
For performance, we use Thumbulator~\cite{thumbulator}, a cycle accurate simulator for the ARM M0+ CPU. 
All our networks use 16-bit fixed point values in Q4.11 format.\footnote{Our solution also applies to networks that use floating point, such as those shown in CSI NN.} 
For fully connected and max pooling layers, we shuffle the order of all loops.
For convolutional layers, we shuffle input channels, the output channels, input rows and input columns. 

\myparagraph{Benchmarks.}
Our benchmarks cover typical networks run on IoT devices.
\textbf{mnist-mlp} and \textbf{mnist-cnn} represent image recognition tasks, which are increasingly popular on IoT devices~\cite{magid2019}.
\textbf{kws-mlp} is a audio keyword spotting network for IoT devices~\cite{zhang2017}.
\textbf{har-cnn} classifies users' activities based on accelerometer data~\cite{ha2016}. 
\textbf{gesture-cnn} takes camera input and classifies the gesture performed to control an IoT system~\cite{zhan2019}. 
\textbf{ecg-ae} uses an autoencoder to detect anomolous readings from ECG data~\cite{Hori2020}.
\textbf{seizure-svm} processes EKG data to identify the on-set of a seizure, so preventing action can be taken~\cite{Seng2012}.

\myparagraph{Software shuffling.}
For all the benchmarks we study, we see that software shuffling adds a significant overhead, up to 271\%.
For MLP networks, shuffling takes longer for larger layers, as the list of indices to be shuffled is longer. 
The overhead for \textit{kws-mlp} is lower compared to the \textit{mnist-mlp} network, as the former has smaller FC layers. 
For the CNN networks, benchmarks with more CONV layers have lower overhead.
There are two reasons for this: 1)~CONV layers have smaller indices which makes shuffling faster and 2)~CONV layers require more computation than FC layers.
For CONV layers, each weight kernel of size $N\times N$ requires $N^2$ multiply accumulate (MAC) operations, while FC require a single MAC operation per weight. 
This higher compute cost amortizes the high cost of software shuffling.
However, networks with fewer CONV layers have very high overhead as the first FC layer has a large number of neurons. 
The overhead from this large FC layer dominates the overhead of software shuffling. 
In contrast, prior work only shows an 18\% overhead for software shuffling~\cite{Brosch2022}.
This low overhead is because they only test a very simple MLP network with 15, 10 and 10 neurons per layer. 
As we evaluate much larger networks, we see significantly higher overheads when using software shuffling.

\myparagraph{Hardware shuffling.}
In contrast to software shuffling, the additional instructions needed for hardware shuffling adds an average of just \latoverhead latency overhead. 
The overhead is higher for the MLP networks as they consist solely of fully connected layers. 
As we shuffle both dimensions (i.e., neurons and weights per neuron) for FC layers, our technique adds more instructions, leading to greater overhead. 
We see lower overhead for CNN networks as they spend more time computing convolutional layers.
Unlike software shuffling, the overhead of our technique does not scale with the size of layers.

\myparagraph{Impact of shuffling on accuracy.}
Shuffling does not affect network accuracy, as all the operations are still performed, merely in a different order.
In contrast, using the weights recovered by the attack when shuffling is used results in a significant loss of accuracy. 
For example, for mnist-mlp the weights recovered with $k=16$, result in an accuracy of just 11.7\%.
It is important to note that shuffling the order of operations does not incur any additional latency due to cache non-locality.
Low-power IoT CPUs such as the ARM M0+ do not use caches.
Thus, all memory accesses take the same number of cycles to complete. 

\subsection{Area, frequency and power analysis}
\label{s:analysis}

As mentioned in Section~\ref{s:solution-hardware}, we opt for a design with $16$ register per set (i.e., bins).
The largest layer in our evaluation is $5760$ neurons.
We therefore use 10-bit registers, allowing us to support a maximum of 16,384 iterations.
With this sizing in mind, we now explore the operating frequency and the area and power overheads of \name.

We design \name in Verilog and synthesize it using the Synopsys Design Compiler Version N-2017.09. 
As IoT devices are typically manufactured using older device technologies~\cite{STMicro-process}, we use the TSMCs 65nm (nominal) process technology. 
For area and delay, we use Cadence Innovus v16.22-s071 and Mentor Graphics ModelSim SE 10.4c.
\name adds just 2.2\%  area to an ARM M0+ SoC manufactured in 65nm~\cite{Myers2015}. 
\name has an $F_{max}$ of 257.83MHz, which is much faster than the clock speed of IoT devices.
Prior works use frequencies ranging from 10MHz to 50Mhz for IoT devices used for ML applications~\cite{Choi2018, Desai2022, Guo2019, Giraldo2019}.
Thus \name has no impact on the $F_{max}$ of the overall system.
We opt to run our CPU at 24MHz, matching prior work~\cite{hicks-isca2017}.
At this frequency, \name incurs a 2.22\% power overhead, compared to a ARM M0+ CPU~\cite{STMicro-power}.
This is in contrast to software shuffling, which, on average, more than doubles the latency and therefore energy cost of computation.

\myparagraph{TRNG.}
We now quantify the randomness required by \name.
Our hardware runs at 24MHz and we use 16 registers per set. 
As we described in Section~\ref{s:solution-latency}, \name produces a new value every 3 cycles.
Therefore, we require $24 \times log_2{(16)} \div 3 = 32$ Mbits/s of randomness.
To satisfy this requirement, we used a TRNG which provides up to 86 Mbits/s of randomness~\cite{Pamula2018}.
The TRNG adds an additional 0.26\% area and 1.06\% power overhead, which brings our total overhead to \areaoverhead area and \pwroverhead power. 

\subsection{Security of shuffling hardware}
\label{s:evaluation-coco}

We now explore whether \name can leak any side channel information that an attacker can use to subvert our solution. 
We use a \textit{formal verification} based approach, which is highly effective in detecting possible side channel leaks.
Formal verification has previously identified leaks in a hardware encryption algorithm, which was previously deemed secure based on attacking captured traces~\cite{arribas2018}. 

We use the \textit{CocoAlma}~\cite{hadzic2021} tool, which takes a Verilog file as input and searches for possible side channel leaks. 
\textit{CocoAlma} checks for any variations in latency or power during operation which could potentially serve as a side channel leak. 
This tool also accounts for hardware leakage effects such as glitches.
We analyze \name using \textit{CocoAlma} and verify that there are no side channel leaks from \name.

\section{Broader applicability of \name}
\label{s:discussion}

In this section, we outline how \name can be used for more than just securing ML algorithms against power-side channel attacks.
First, we describe two security-critical applications that can be secured using \name.
We then provide an overview of other types of attacks against neural networks running on IoT devices and describe how \name can also effectively prevent these attacks. 

\subsection{Other applications}
\label{s:discussion-otherapps}

\myparagraph{Elliptical curve cryptography (ECC).}
ECC is a public-key cryptography scheme based on elliptic curves over finite fields~\cite{Bruen2021}.
ECC encodes keys as coefficients of polynomials.
Prior work shows that ECC leaks side channel information, which can be used to recover private keys~\cite{Danger2013}.
Attacks target the \textit{elliptic curve multiplication (ECM)} operation, commonly implemented using the `double-and-add' method~\cite{Jiang2016}.
ECM takes a point $p$ as input and loops over each bit of $p$; if the bit is $1$, ECM performs an \textit{add} operation.
Thus, iterations which take longer have a $1$ in that bit position. 
With \name, we can shuffle the order in which bits are accessed each time, which prevents the attacker from learning which bits are $1$.
As ECC uses at least $224$ bit keys~\cite{Chen2019}, shuffling increases the number of possible permutations tremendously.

\myparagraph{Biometric authentication.}
An emerging use case for IoT devices is for biometric authentication~\cite{Golec2020}.
An example of this is a fingerprint recognition system, such as those commonly used in laptops.
Prior work shows that such systems are susceptible to side channel attacks~\cite{Galbally2020}.
Specifically, the CPA attack (outlined in Section~\ref{s:attack}) can be used to learn each user's stored fingerprint data~\cite{Chouta2014}.
The recognition system is implemented as a set of nested for loops, which can be shuffled using \name to obscure this side channel.

\subsection{Other attacks}
\label{s:evaluation-otherattacks}

\myparagraph{Floating point timing attack.} 
The difference in time taken by floating point multiplication based on the input values~\cite{Gongye2020} can be used to mount an attack.
In the IEEE-754 32-bit floating point format, the smallest number using the normal representation is $1.0\times2^{-126}$.
Numbers smaller than this are called subnormal; operations involving subnormal numbers take much longer than operations using only normal numbers. 
For example, on an x86 system, $(normal\times normal=subnormal)$ takes 124 cycles, while $normal\times normal=normal$ takes only 10 cycles.  
During network inference, each $(input\times weight)$ operation has a specific $input$ value which will cause the output to become subnormal.
The attacker sweeps the $input$ to find this threshold value and then uses that to learn the $weight$.
The attacker can then recover all the weights of the first layer and repeat the process for the other layers.
While this attack is limited to networks that use floating point numbers, it requires less equipment as it relies on timing rather than power.
However, this attack still requires each operation to occur in the same place in each trace, so the attacker can try multiple $input$ values to find the threshold $input$ value.
\name prevents this attack by randomizing the order of operations, and preventing the iterative search.

\myparagraph{Fault injection attacks.}
The attacks discussed thus far have focused on stealing the model; in contrast, fault injection attacks cause the model to operate in an abnormal way~\cite{Breier2018, Breier2021}.
For example, in the network used for `chip-and-pin', fault injection can be used to classify a fraudulent transaction as legitimate.
Attackers inject `faults' into the system while it is running the model, forcing it to mis-classify its inputs.
Prior work shows a practical attack using lasers to inject faults~\cite{Breier2018}.
To counteract such attacks, techniques have been proposed to detect faults~\cite{Javaheripi2021, Wilson2020}. 
However, detection techniques incur high overheads and are not 100\% accurate.
To minimize the chance of detection, the attacker must inject as few faults as possible~\cite{Zhao2019, Ghavami2021}.
Prior work shows that a mis-classification can be forced with just 4 injected faults~\cite{Ghavami2022}.
However, the attacker must have full knowledge of the model, to determine the exact points where faults must be injected.
By shuffling the order of operations, \name prevents the attacker from determining the exact location for fault injection. 
The attacker must therefore inject many more faults, and therefore significantly increase the chances of detection. 

\section{Related work}
\label{s:related}

In this section, we present some related work on \textit{masking}, which is another commonly used technique for preventing side-channel attacks.
We also list prior works on securing machine learning in a broader context.

\subsection{Shuffling}
\label{s:related-shuffling}

Shuffling was first proposed as a technique to secure AES encryption against side channel attacks~\cite{mangard2007}.
Most shuffling techniques target the $16$ S-Box operations performed in AES~\cite{veyrat2012}. 
We now detail prior work which perform shuffling in software and hardware.

\myparagraph{Software.}
One approach to shuffle the order of operations is to pick a random index to start at each time. 
As this only requires calculating one random value, it adds significantly less overhead~\cite{feldhofer2008power, medwed2010fresh, Moradi2011}.
However, follow-on work shows that this approach does not significantly improve security as it only results in $N$ permutations instead of $N!$~\cite{veyrat2012}.
Another approach is to combine shuffling with inserting dummy instructions to further mis-align the recorded power traces~\cite{lee2020}.
However, this approach is challenging as the dummy instructions must appear genuine to the attacker or else they can easily remove them from the trace before analysis.

Other approaches perform `fully shuffling', for securing the S-Box operation of AES running on a low-power CPU~\cite{Bayrak2012}.
This is done by unrolling the loop which computes the $16$ S-Box computations and running these steps in a random order.
This technique would be impractical for neural networks due to the larger, arbitrary number of neurons and weights in neural networks.

\myparagraph{Hardware.}
Shuffling in hardware has also been implemented for AES on FPGA~\cite{Wang2013, patranabis2016, Dhanuskodi2019}.
Techniques that combine hardware and software to perform AES shuffling have also been proposed~\cite{ge2018}.
Adding hardware for shuffling has also been proposed for other encryption algorithms such as elliptic curve cryptography~\cite{Chen2019} and lattice-based cryptography~\cite{Chen2023}. 
These approaches are also restricted to shuffling $2^N$ iterations, while \name supports shuffling any number of iterations.
Shuffler~\cite{Shuffler} and Morpheus~\cite{Gallagher2019} employ shuffling to protect against code reuse attacks, while we defend against side channel attacks.

\myparagraph{Shuffling for NNs.}
Dubey et al. add shuffling to an accelerator for binary neural networks, to defend against side-channel attacks~\cite{Dubey2022}.
However, they only shuffle the starting index which leads to a significantly smaller number of permutations.
They explicitly state that they do not full shuffling as the values are not powers-of-2, which is the problem solved by our approach.

\subsection{Masking}
\label{s:related-masking}

One popular technique to obfuscate side-channels is to \textit{mask} secret data by \textit{splitting} this data into several parts and operating on each part separately. 
Mathematically, the secret information $s$ is split into $d$ parts $s_1, s_2, ... s_d$ such that $s_1\oplus, s_2\oplus...\oplus s_d = s$. 
The masking must be done such that any subset of less than $d$ shares are statistically independent of $s$. 
A simple way to achieve this is by picking $s_1, ..., s_{d-1}$ uniformly at random (the masks), and setting $s_d = s \oplus s_1 \oplus ... \oplus s_{d-1}$ (the masked variable). 
The masking is then said to be of order $d-1$. 
However, a $(d-1)^{th}$ order masked implementation is susceptible to a $d^{th}$ order attack, which analyses all $d-1$ shares collectively to recover secret information.

Masking has been extensively studied for securing encryption algorithms such as AES~\cite{Kim2011}, Saber~\cite{Kundu2022} and Midori64\cite{Ghanbari2022}.
Masking has also been applied to CPUs to prevent side-channel attacks but incur a $141\times$ latency overhead~\cite{Antognazza2021}.
Similar to our approach, Dubey et al. modify a RISC-V CPU to mask operations during network inference, but add $2\times$ latency overhead~\cite{Dubey2023}, compared to just \latoverhead added by \name.

Techniques to secure neural networks accelerators by masking have also been proposed, although these techniques impose significant latency (up to 2.8$\times$) and area (up to 5.9$\times$) overheads~\cite{maskednet, bomanet}.
Maji et al. propose a masking-based neural network accelerator to prevent power side-channel attacks, which adds $1.4\times$ latency and $1.64\times$ area overhead and only targets fully connected layers~\cite{Maji2023}.
In contrast, \name adds just \latoverhead latency and \areaoverhead area overhead to an ARM M0+ SoC.

\subsection{Machine learning security.} 
Prior work has shown attacks to steal networks deployed on cloud service such as AWS or Google Cloud~\cite{Shokri2017, tramr2016stealing,fredrikson2014}.
These attacks require the attacker to repeatedly query an online network to train their own network to match accuracies. 
However, unlike the attacks we counter, these attacks require access to the same training data as the online model. 

Attacks against ML algorithms using cache side channels have also been proposed~\cite{hong2018, telepathy2020}.
These attacks leverage the difference in cache access timing to infer information.  
However, as IoT devices typically lack caches, such attacks do not apply to them.
Similarly, the memory access pattern of neural network accelerators has also been used as a side-channel to recover information~\cite{weizhe2018}.
The authors are able to reverse engineer the network architecture from the memory access patterns observed during inference. 
This attack does not apply to low-power embedded systems, where all memory accesses take the same number of cycles. 


Another solution to obscure side channel leakage due to memory access patterns is oblivious RAM (ORAM)~\cite{Goldreich1987}. 
ORAM however cannot be used to hide power side channels, which is the focus of our work~\cite{Costa2017}.
Also, ORAM imposes a large 100$\times$ overhead, compared to just a few percent for our technique~\cite{aga2017}.

\section{Conclusion}

We show that shuffling is an effective technique to prevent side channel attacks against neural networks running on IoT devices.
We detail a new attack against software shuffling -- proposed in prior work -- leaks information which can be used to obviate the security benefits of shuffling.
To perform secure shuffling, we propose \name, hardware added as a functional unit within the CPU. 
\name uses a novel counter-based approach, to effectively shuffle the large, arbitrary sizes of neural network layers.
\name adds just \latoverhead latency overhead, compared to over 100\% in the case of software shuffling.
We show that \name effectively secures the weights of neural networks and can also be used to secure other applications.
We also describe how \name is effective at preventing other side channel attacks such as floating point timing attacks and fault injection attacks.
\name adds just \areaoverhead area and \pwroverhead power overhead, without itself leaking any side channel information.

\section*{Acknowledgements}
We thank the members of NEJ group for their valuable feedback.
We acknowledge the support of the Natural Sciences and Engineering Research Council of Canada (NSERC) Discovery Grant RGPIN-2020-04179. 
This research was undertaken, in part, thanks to funding from the Canada Research Chairs program.


\bibliographystyle{IEEEtranS}
\bibliography{ref}
	
\end{document}